\documentstyle[11pt,epsf]{article}
\begin{document}
\bibliographystyle{unsrt}
\input{rotate}
\textwidth=18.5truecm
\textheight=23.0truecm
\hoffset=-1.0truecm
\voffset=-2.1truecm
\newcommand{\bea}{\begin{eqnarray}}
\newcommand{\eea}{\end{eqnarray}}
\newcommand{\be}{\begin{equation}}
\newcommand{\ee}{\end{equation}}
%
%
%
\def\shiftleft#1{#1\llap{#1\hskip 0.04em}}
\def\shiftdown#1{#1\llap{\lower.04ex\hbox{#1}}}
\def\thick#1{\shiftdown{\shiftleft{#1}}}
\def\b#1{\thick{\hbox{$#1$}}}

\title{Intrinsic quadrupole moment of the nucleon}

\author{
A. J. Buchmann$^{1}$\thanks{email:alfons.buchmann@uni-tuebingen.de}, 
E. M. Henley$^{2}$\thanks{email:henley@phys.washington.edu} \\
Institut f\"ur Theoretische Physik, Universit\"at T\"ubingen\\
Auf der Morgenstelle 14, D-72076 T\"ubingen, Germany \\
$^2$ Department of Physics and Institute for Nuclear Theory, Box 351560, \\ 
University of Washington, Seattle, WA 98195, U.S.A.\\ }

\maketitle

\bigskip
\medskip

\begin{abstract}
We address the question of the intrinsic quadrupole moment $Q_0$ of 
the nucleon in various models.
All models give a positive intrinsic quadrupole moment for the proton.
This corresponds to a prolate deformation.
We also calculate the intrinsic quadrupole 
moment of the $\Delta(1232)$. All our models lead to a negative
intrinsic quadrupole moment of the $\Delta^+$ corresponding
to an oblate deformation.

\end{abstract}

\centerline{ PACS numbers: 12.39Jh,  14.20Dh, 14.20Gk, 13.40Em}

\section{\bf Introduction}

Electron-proton scattering and atomic Lamb shift 
measurements have shown that
the spatial extension of the proton charge distribution (charge radius) 
is about $r_p \approx 0.9$ fm \cite{Ros00}. 
In addition to the charge radius, the elastic electron scattering data  
provide precise information on the radial charge density 
$\rho(r)$ of the proton. 
However, they do not allow to draw any 
definite conclusions concerning possible deviations of the proton's
shape from spherical symmetry.

In order to learn something about the shape of a spatially extended 
particle one has to determine its {\it intrinsic} quadrupole moment 
\cite{Boh75} 
\be 
Q_0=\int d^3r \rho({\bf r}) (3 z^2 - r^2), 
\ee
which is defined with respect to the body-fixed frame. 
If the charge density is concentrated along the $z$-direction 
(symmetry axis of the particle),  
the term proportional to $3z^2$ dominates, $Q_0$ 
is positive, and the particle is prolate (cigar-shaped).
If the charge density is concentrated in the equatorial plane perpendicular
to $z$, the term proportional to $r^2$ prevails, $Q_0$
is negative, and the particle is oblate (pancake-shaped).
The intrinsic quadrupole moment $Q_0$ must be distinguished
from the {\it spectroscopic} quadrupole moment $Q$ measured in 
the laboratory frame. Due to angular momentum selection rules, a spin $J=1/2$  
nucleus, such as the nucleon, does not have a spectroscopic 
quadrupole moment; however, it
may have an {\it intrinsic} quadrupole moment as was realized 
more than 50 years ago \cite{Boh75}.
Some information on the shape of the nucleon or any
other member of  the baryon octet can be obtained 
by electromagnetically exciting the baryon 
to spin $J=3/2$ or higher spin states.

With the Laser Electron Gamma Source at Brookhaven and various
continuous electron beam accelerators,   
one can carry out high precision pion production experiments on 
the nucleon. In these experiments
a photon (real or virtual) excites the $N(939)$ to, for example, 
the $\Delta$(1232) resonance with spin $J=3/2$. Magnetic dipole $(M1)$,
electric quadrupole 
($E2$) and/or charge quadrupole ($C2$) excitation modes are allowed
by angular momentum conservation and invariance under the parity
transformation.
Once produced, the $\Delta$(1232) hadronically 
decays to a nucleon and pion. By measuring the momenta of the 
final state nucleon and pion in coincidence, individual 
electromagnetic multipoles can be extracted.

The $E2$ and $C2$
multipoles carry the information about the intrinsic deformation
of the nucleon. 
If the charge distribution 
of the initial and final three-quark states were spherically symmetric,
the $E2$ and $C2$ amplitudes would be zero (Becchi-Morpurgo 
selection rule \cite{Bec65}).
The experimental values for these quadrupole amplitudes are small compared to
the dominant magnetic dipole transition, but they are clearly nonzero. 
Recent data \cite{Bec97,Bla97,Kal97}
indicate that the ratio of the electric quadrupole amplitude to the 
magnetic dipole amplitude is at least 
$E2/M1 \approx -3\%$.
A $C2/M1$ ratio of the same sign
and comparable magnitude has been measured \cite{Bar98}.
Recently, an 
experimental value for the $N \to \Delta$ quadrupole transition moment
has been derived 
$Q_{exp}^{N \to \Delta} = -0.108 \pm 0.009 \pm 0.034$ fm$^2$
\cite{Bla97}.
From these measurements one can conclude that the 
nucleon and the $\Delta$ are intrinsically deformed. 
However, the magnitude and sign of the intrinsic $N$ and $\Delta$ 
deformation can only be calculated within a model.

There is considerable 
uncertainty in the literature concerning the implications of the experimental
$C2/M1$ and $E2/M1$ ratios for the intrinsic 
deformation of the nucleon. Even with respect to the sign of the intrinsic 
nucleon deformation there is no consensus. For example, references 
\cite{Gia79,Ven81,Ma83} conclude that the nucleon is oblate, while
references \cite{Cle84,Mig87,Buc99} find a prolate nucleon deformation.
Several authors speak only about `deformation' without 
specifying the sign. 

The purpose of this paper is to calculate the intrinsic quadrupole
moment of the proton $Q_0^p$ in various models. In particular,
we want to predict its sign, i.e., we want to find out 
whether the proton is prolate or 
oblate. Before doing this, we discuss the possible origins of nucleon 
deformation in the quark model.

\section{Sources of quadrupole deformation } 

\begin{figure}[htb]
$$\mbox{
\epsfxsize 10.5 true cm
\epsfysize 3.0 true cm
\setbox0= \vbox{
\hbox { \centerline{
\epsfbox{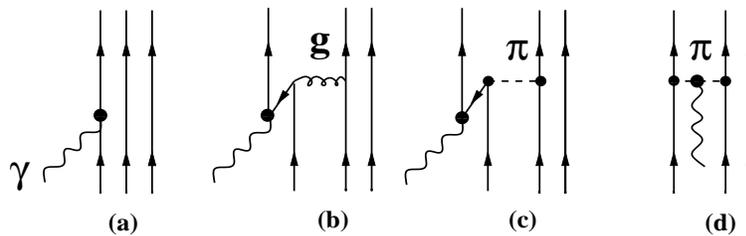}
}} 
} 
\box0
} $$
\vspace{-0.5cm}
\caption[Exchange currents]{Feynman diagrams of the four vector
current $J^{\mu}=(\rho, {\bf J})$:
(a) one-body current ${\bf J}^{\mu}_{[1]}$, 
and (b-d) two-body gluon and pion exchange currents 
${\bf J}^{\mu}_{[2]}$. The one-body charge density leads to  
a quadrupole operator of the type given in Eq.(\ref{onebody}).
Both gluon and pion exchange charge densities 
produce a two-body quadrupole operator of the type given in 
Eq.(\ref{decomp}). }
\label{feynmec}
\end{figure}

Two different sources contribute to a quadrupole deformation 
of baryons. First, tensor forces between quarks 
lead to $D$-state admixtures in the single-quark wave
functions of a baryon, and consequently to a deviation of the 
{\it valence quark} distribution from spherical symmetry. 
The one-gluon exchange interaction was originally proposed 
to provide the required tensor force \cite{Ger82,Isg82}. 
An external photon can induce a quadrupole transition, for example, by 
lifting an $S$ state quark in the $N$ into a $D$ state 
in the $\Delta$ via the one-body 
current in Fig. \ref{feynmec}(a). 

Second, quark-antiquark pairs and gluons are present in a 
physical baryon. These degrees of freedom also 
contribute to the observed quadrupole transition.
We refer to the $q{\bar q}$ and gluon degrees of
freedom generically as {\it nonvalence quark degrees of freedom}. 
The latter are effectively described as spin-dependent 
two-quark operators in the electromagnetic current \cite{Buc91}.
The two-body terms in the charge and current operators shown in 
Fig. \ref{feynmec}(b-d) 
arise as a result of eliminating the $q {\bar q}$ and gluon 
degrees of freedom from the wave function, very much in the same way
as the two-body potentials in the Hamiltonian result from the elimination
of the ``exchange particle'' degrees of freedom from Hilbert space.

Hence, tensor force induced $D$-waves in the single-quark wave function
(one-quark quadrupole transition) {\it and} nonvalence quark degrees of 
freedom  (two-quark quadrupole transition) contribute to the deformation 
of baryons. In principle, there can also be three-quark operators. 
We neglect them here. In different paper \cite{Buc00} 
we argue that their contribution is suppressed by at least $1/N_c$ 
compared to the two-body terms. 

Experimental amplitudes contain both mechanisms, 
so that one cannot readily distinguish 
between the two different excitation modes (one-quark vs. two-quark currents).
If, however, the single quark transition mode is strongly suppressed
\cite{Bec65}, and if one measures an $E2$ strength
that is large compared to the single quark estimate, one may conclude
that the deformation resides in the nonvalence quark degrees of freedom,
effectively described by the two-body current operators.

\subsection{Single quark operator: Deformed valence quark orbits }

In a multipole expansion of the one-body charge operator,
the term proportional to the spherical harmonic of rank two,
$Y^2({\hat{\bf r}_i})$ provides the one-body quadrupole operator:
\be 
\label{onebody}
{\hat Q}_{[1]} = \sqrt{\frac{16 \pi}{5}} \sum_{i=1}^3 
e_i r_i^2 Y^2_0({\bf r}_i)
               = \sum_i e_i (3 z_i^2 - r_i^2), 
\ee
where the sum is over the three quarks in the baryon.
Obviously, this one-body operator 
needs $D$ waves in the nucleon or $\Delta$ in order to make a 
nonvanishing contribution. In a two-state model (only S and D waves), 
the nucleon and $\Delta$ wave functions can be written as 
\begin{eqnarray}
\label{wave}
\left \vert N \right \rangle & = &
a_S \left \vert (S=1/2 \,, L=0) J=1/2 \right \rangle +
a_D \left \vert (S=3/2 \,, L=2) J=1/2 \right \rangle  \nonumber \\
\left \vert \Delta \right \rangle & = &
b_S \left \vert (S=3/2 \,, L=0) J=3/2 \right \rangle +
b_D \left \vert (S=1/2 \,, L=2) J=3/2 \right \rangle ,
\end{eqnarray}
where the quark spin $S$ couples with the
orbital angular momentum $L$ to the total angular momentum $J$ 
of the baryon. 
The $D$-states in Eq.(\ref{wave}) are of mixed symmetry type
with respect to the exchange of quarks 1 and 2.
We have purposely omitted the symmetric $D$-state in the $\Delta$
wave function $c_D \vert (S=3/2, L=2) J=3/2 \rangle$, 
which is smaller in magnitude than the one listed here.
The negative relative sign of the $D$ wave amplitude $a_D=-0.04$ 
\cite{Ger82,Isg82} with respect to 
the $S$ wave amplitude  indicates an oblate deformation of the 
valence quark distribution in the nucleon. Similarly, the positive sign  
of the $D$ state amplitude $b_D=0.07$ in the $\Delta$ corresponds
to a prolate deformation of its valence quark distribution.

\begin{figure}[htb]
$$\mbox{
\epsfxsize 10.0 true cm
\epsfysize 7.0 true cm
\setbox0= \vbox{
\hbox { \centerline{
\epsfbox{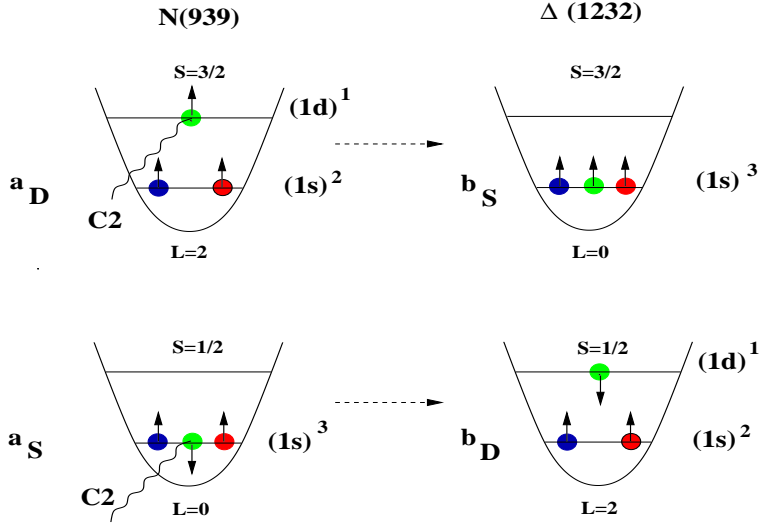}
}  } 
} 
\box0
} $$
\vspace{-0.5cm}
\caption[shell]{
$N\to \Delta$ quadrupole transition via the one-body
quadrupole operator ${\hat Q}_{[1]}$ of Eq.(\ref{onebody}) 
coming from the one-quark current in Fig.\ref{feynmec}(a).
In this single-quark transition, the absorption of a $C2$ photon is 
only possible if either the nucleon (left) or the $\Delta$ (right) 
contains a $D$ wave admixture (deformed valence quark orbit). 
Actually, the $ N \to \Delta $ 
quadrupole transition  due to ${\hat Q}_{[1]}$ is a coherent 
superposition of the two orbital angular momentum 
changing but intrinsic spin $S$ conserving one-body 
transitions (upper and lower part of the figure).
The resulting quadrupole transition matrix element   
of Eq.(\ref{c2imp}) 
is suppressed due to the small $D$ wave admixtures 
in the $N$ und $\Delta$ wave functions. }
\label{shellmodel}
\end{figure}

Applied to the $N \to \Delta$ quadrupole transition,
the one-body quadrupole operator 
${\hat Q}_{[1]}$ sandwiched between  
the $N$ and $\Delta$ wave functions gives for the 
quadrupole transition moment \cite{Gia90}
\begin{equation}
Q_{p \to \Delta^+}  =  - b^2 {4 \over \sqrt{30}} 
\left ( a_S b_D - a_D b_S \right ),
\label{c2imp}
\end{equation}
where the small $a_Db_D$ term has been neglected. 
Here, the harmonic oscillator parameter $b$ 
describes the spatial extension of the baryon wave function,
and we refer to it as quark core (matter) radius. 
The two terms of this single quark current matrix element are 
schematically shown in Fig. \ref{shellmodel}.

Using standard $D$-state admixtures \cite{Ger82,Isg82} 
and an unphysically large quark core radius of $b=1$ fm in 
Eq.(\ref{c2imp}) one could describe the experimental transition quadrupole 
$Q^{exp}_{p \to \Delta^+} \approx -0.11$ fm$^2$ \cite{Bla97} 
by the one-body term alone. However, from the description of 
the baryon spectrum it is known that one needs an average single-quark 
excitation energy $\omega=1/(m_q b^2) \approx 500$ MeV \cite{Gia90}, 
which implies a quark core radius $b \approx 0.5 $ fm. 
With this smaller value for the quark matter radius $b$ 
one obtains only $20\%$ of the 
experimental quadrupole strength.

That the single-quark excitation is not the dominant
quadrupole excitation mechanism becomes particularly 
apparent when one considers the $E2/M1$ ratio in the electromagnetic 
excitation of the $\Delta^+$. Calculations of the $E2/M1$ ratio based 
on spatial single quark currents give $E2/M1=-0.1\%$ \cite{Cap90}, which 
is an order of magnitude smaller than recent experiments. 
Thus, the experimental $E2/M1\approx -3\%$ ratio \cite{Bec97,Bla97} 
cannot be solely described by a single-quark transition. 
Other degrees of freedom must be taken into account. These points
have recently been discussed in more detail \cite{Buc97}.

\subsection{Two-quark operator: Deformed $q\bar q$ cloud}

\begin{figure}[htb]
$$\mbox{
\epsfxsize 10.0 true cm
\epsfysize 4.5 true cm
\setbox0= \vbox{
\hbox{ 
\epsfbox{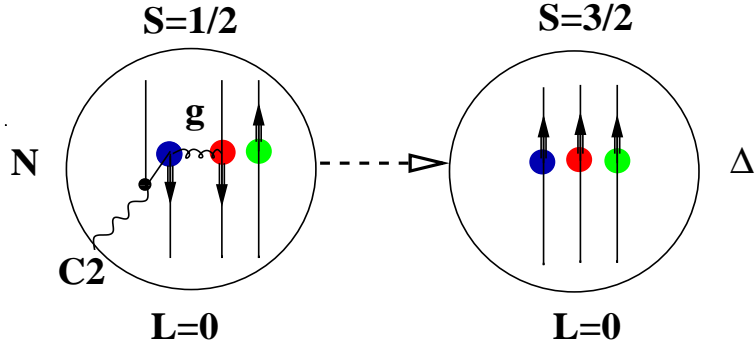}
} 
} 
\box0
} $$
\vspace{-0.3cm}
\caption[Spinflip]{
$N\to \Delta$ quadrupole transition via the two-body
quadrupole operator ${\hat Q}_{[2]}$ of Eq.(\ref{decomp}) 
originating from, e.g., the two-body gluon exchange current 
in Fig.\ref{feynmec}(b). 
The quadrupole transition proceeds by absorbing a $C2$ photon 
on a quark-antiquark pair with all valence quarks remaining in an 
$S$ state. 
This double spin flip quadrupole amplitude describes the deformation
of the $q\bar q$ cloud in the nucleon. It 
can be parameter-independently expressed
in terms of the neutron charge radius 
$Q_{p \to \Delta^+} = r_n^2/\sqrt{2} =-0.80$ 
fm$^2$ \cite{Buc97}. 
This prediction is in agreement with the recent extraction of 
$Q_{p \to \Delta}^{exp}=-0.108\pm 0.009 \pm 0.034$ by the LEGS group 
\cite{Bla97}
indicating that the major part of the quadrupole strength resides 
in the collective $q \bar q$ degrees of freedom.}
\label{flip2}
\end{figure}

Many people believe that the valence quarks must move 
in $D$ waves in order to obtain a nonvanishing quadrupole
moment. However, a two-body spin tensor in the charge 
operator also generates a quadrupole moment even when
the valence quarks are in pure 
$S$ states \cite{Buc97}. A similar observation was also made by Morpurgo
\cite{Mor89}. 
The two-body quadrupole operator generated by the $q \bar q$ pair currents
of Fig. \ref{feynmec} (b-c) 
\begin{equation}
\label{decomp} 
{\hat Q}_{[2]} = B
\sum_{i\ne j=1}^3 e_i  \left ( 3 \sigma_{i \, z}  \sigma_{j\,  z} - 
{\b{\sigma}}_i \cdot {\b{\sigma}}_j \right ) 
\end{equation} 
acts in spin and isospin space, whereas the one-body operator 
in Eq.(\ref{onebody}) acts in isospin and orbital space. 
The constant $B$ with dimension fm$^2$ 
contains the orbital and color matrix elements.
As a spin-tensor of rank 2, the operator ${\hat Q}_{[2]}$  
may simultaneously flip the spin of two quarks (double spin flip) 
in such a way that the total 
spin changes from 1/2 to 3/2 
(see Fig. \ref{flip2})\footnote{
The name double spin flip becomes clear  if one rewrites one term 
in Eq.(\ref{decomp})
in the spherical basis with spin raising ($\sigma_+$) and 
lowering ($\sigma_-$) operators, e.g., 
$B\,  e_1 ( 2 \sigma_{1\, z} \sigma_{2\, z}
+ \sigma_{1 \, - } \sigma_{2 \, + }   + \sigma_{1 \, + } \sigma_{2 \, - })$.}.
  
We emphasize that {\it although the operator ${\hat Q}_{[2]}$  
formally acts on valence quark spin
states, it does  not describe the deformation of the valence quark core.
Instead, it reflects that the physical nucleon 
contains $ q \bar q$ sea-quarks whose distribution 
deviates from spherical symmetry.}

Using a quark model with two-body exchange currents the constant $B$ 
has been calculated.
It was found that 
the $N \to \Delta$ and $\Delta$ quadrupole moments can be expressed 
in terms of the neutron charge radius
$r_n^2$ as follows \cite{Buc97}
\begin{eqnarray}
\label{relations}
\sqrt{2} \, Q_{p \to \Delta^+} & = & 
Q_{\Delta^+} = 4 B = r_n^2.
\end{eqnarray}
The reason for the existence of such a relation between 
$Q_{\Delta}$ and $r_n^2$ is that
both observables are dominated by exchange currents.
When expanding the gluon and pion  exchange charge operators 
in Fig.\ref{feynmec}(b-c) into Coulomb multipole operators 
one finds a fixed relative strength between the monopole term
$C0 = -2\, B  \sum e_i {\b{\sigma}}_i \cdot {\b{\sigma}}_j$ 
(giving rise  to a nonvanishing 
neutron charge radius \cite{Buc91}) and the quadrupole term
$C2 = B \sum e_i 
(3 \, \sigma_{i\, z} \, \sigma_{j\, z} - \b{\sigma}_i \cdot \b{\sigma}_j)$   
(leading to a nonzero transition quadrupole moment \cite{Buc97}).   
As a result, one obtains the same
analytic expression for $r_n^2$ 
and $\sqrt{2} Q_{p \to \Delta^+}$, suggesting  
that the deformation of the nucleon is closely connected
to the nonvanishing neutron charge radius.
The quadrupole moment calculated from the experimental neutron charge radius 
according to Eq.(\ref{relations}) is 
in good agreement with the transition quadrupole moments extracted 
from the $E2/M1$ and $C2/M1$ measurements \cite{Bla97}. 

Let us summarize. The quark model shows that {\it  both} 
the $N(939)$ ground state and the excited $\Delta(1232)$ state
are deformed. This conclusion is reached in 
the single quark transition model where tensor forces lead to 
$D$ state admixtures in {\it both} the $N$ and $\Delta$ wave function.
It is also obtained in the two-quark transition model,  where 
exchange currents produce a nonzero  quadrupole moment of the $S=1$ diquark 
in the $N$ and $\Delta$. It is impossible to 
have a spherical nucleon and a deformed $\Delta$ or vice versa.
If the quarks interact via vector or pseudoscalar type potentials,
either both baryons are deformed or both are spherical. 
The latter possibility is ruled out by experiment. 
Furthermore, we understand
that two-body $q \bar q$ pair currents provide the major contribution
to baryon quadrupole moment, suggesting that the deformation
resides in the  baryon's $q\bar q$ cloud.

\section{Intrinsic deformation of the nucleon}

In this section we discuss the intrinsic quadrupole moment 
of the nucleon. 
Here, {\it intrinsic} quadrupole moment means the one obtained 
in a body-fixed coordinate system that rotates with the nucleon.
Most work that
deals with the problem of nucleon deformation in the 
quark model does not distinguish between intrinsic and 
spectroscopic (measured) quadrupole moment.
This is all the more surprising 
since the shape of the nucleon is in the first place 
related to the intrinsic, and not to the spectroscopic 
quadrupole moment \cite{Boh75}.

The spectroscopic quadrupole moment of the nucleon is zero.
Nevertheless, the nucleon can have an {\it intrinsic} 
quadrupole moment. This is analogous to 
a deformed $J=0$ nucleus. All orientations of a deformed $J=0$ 
nucleus are equally probable, which
results in a spherical charge distribution
in the ground state and a vanishing quadrupole moment $Q$ in the
laboratory. The intrinsic quadrupole moment 
$Q_0$ can then only be obtained  
by measuring electromagnetic quadrupole transitions between the ground and 
excited states, or by measuring the quadrupole moment of an excited state 
with $J > 1/2$ of that nucleus. 
If a sufficient number of quadrupole transitions to excited states
are known,
the intrinsic quadrupole moment could be extracted from the data
in a model-independent way as suggested by Kumar \cite{Kum72}. Here, we 
follow a somewhat different approach, which is less general 
than Kumar's method. 

Given only the experimental information 
of the $\Delta$ and the $N \to \Delta$ quadrupole moments what
can we learn about the intrinsic quadrupole moment of the nucleon?
The answer which we give is model-dependent. However, with respect 
to the sign of the intrinsic quadrupole moment all models 
studied in this paper yield 
the same answer, namely that the nucleon is prolate-shaped. 
In the following three sections, we will use three different models 
of the nucleon and $\Delta$, 
(i) a quark model, (ii) a collective model, (iii) and a pion cloud model
in order to calculate the intrinsic quadrupole moment of the nucleon.

\subsection{Quark model}

In standard notation the $SU(4)$ spin-flavor part of the proton 
wave function is composed of a spin-singlet and a spin-triplet
part
\begin{eqnarray}
\label{protonwave}
\vert p \rangle & = & {1 \over \sqrt{2} }
\biggl \lbrace {1 \over \sqrt{6}}
 \vert \left ( 2uud - udu -duu \right ) \rangle 
{1 \over \sqrt{6}}   
\left ( 2 \uparrow \uparrow \downarrow - \uparrow \downarrow \uparrow
- \downarrow \uparrow \uparrow \right ) \rangle \nonumber \\
& & + {1 \over \sqrt{2}} \vert 
\left ( udu -duu \right ) \rangle  \vert
{1 \over \sqrt{2}} 
\left ( \uparrow \downarrow \uparrow - \downarrow \uparrow \uparrow \right )
\rangle  
\biggr \rbrace .
\end{eqnarray}
The angular momentum coupling factors $2$, $-1$, $-1$ in front of the
three terms in the spin symmetric proton wave function 
express (i) the coupling of the first two quarks 
to an $S=1$ diquark, and (ii) the coupling of the $S=1$ diquark with 
the third quark to total $J=1/2$. 

Sandwiching the quadrupole operator ${\hat Q}_{[2]}$ 
between the proton's spin-flavor wave function 
yields a vanishing spectroscopic quadrupole moment.
The reason is clear. The spin tensor ${\hat Q}_{[2]}$ applied 
to the spin singlet wave function gives zero, and  when acting on 
the proton's spin triplet wave function it gives 
\begin{eqnarray}  
\label{intquark1}
\left ( 3 \sigma_{1 \, z} \sigma_{2 \, z} - 
\b{\sigma}_1 \cdot \b{\sigma}_2 \right ) 
{1 \over \sqrt{6}} \left \vert 
\left ( 2 \uparrow \uparrow \downarrow - \uparrow \downarrow \uparrow
- \downarrow \uparrow \uparrow \right ) \right \rangle  
& = & {4 \over \sqrt{6} } 
\vert \left (\uparrow \uparrow \downarrow + \uparrow \downarrow \uparrow
+ \downarrow \uparrow \uparrow \right ) \rangle,
\end{eqnarray} 
where the right-hand side is a spin 3/2 wave function, 
which has zero overlap with the spin 1/2 wave function of the proton 
in the final state. Consequently, 
the spectroscopic quadrupole moment 
\be
Q_p = \langle p \vert \hat Q_{[2]} \vert p \rangle = 
B \left ( 2 - 1 -1 \right ) = 0
\ee
vanishes due to 
the spin coupling coefficients in  $\vert  p \rangle $.

Although the spin $S=1$ diquarks ($uu$ and $ud$) in the proton have
nonvanishing quadrupole moments, the angular momentum 
coupling of the diquark spin to the spin of the third quark 
prevents this quadrupole moment from being observed. Setting ``by hand'' 
all Clebsch-Gordan coefficients in the spin part of 
the proton wave function of Eq.(\ref{protonwave}) equal to 1, 
while preserving the normalization, one obtains  
a modified ``proton'' wave function $\vert {\tilde p} \rangle $
\begin{eqnarray}  
\label{intquark2}
\vert {\tilde p} \rangle & = & 
 {1 \over \sqrt{2} } \biggl \lbrace 
\left \lbrack   \vert 
{1 \over \sqrt{6}}\left ( 2uud - udu -duu \right ) \rangle + 
 {1 \over \sqrt{2}} 
\vert \left ( udu -duu \right ) \rangle \right \rbrack  \nonumber \\
& & 
{1 \over \sqrt{3}} \vert 
\left (\uparrow \uparrow \downarrow +\uparrow \downarrow \uparrow
+\downarrow \uparrow \uparrow \right ) \rangle \biggr \rbrace. 
\end{eqnarray} 
The renormalization of the Clebsch-Gordan coefficients is undoing the 
averaging over all spin directions, which renders the intrinsic
quadrupole moment unobservable.
Note that we do not modify the flavor part of the wave function 
in order to ensure that we deal with a proton\footnote{In the case of the 
neutron we must divide by the negative charge of the $dd$ 
diquark. We then obtain $Q_0^n=Q_0^p$.}.

We consider the expectation value of the two-body quadrupole operator
${\hat Q}_{[2]}$ in the state of the spin-renormalized proton wave function
$\vert \tilde p \rangle $  
 as an estimate 
of the {\it intrinsic} quadrupole moment of the proton $Q_0^p$
\begin{equation}
\label{int1}
Q_0^p = \langle {\tilde p} \vert Q_{[2]} \vert {\tilde p} \rangle
=2 B \left (  \frac{2}{3} -\frac{8}{3} \right ) = - 4 B = - r_n^2,  
\end{equation}
where the two contributions arise from the spin 1 diquark with projection
$M=1$ and $M=0$. The latter dominates. 
Comparing with Eq.(\ref{relations}), 
we find that the {\it intrinsic} quadrupole moment of the proton is equal to 
the {\it negative} of the neutron charge radius $r_n^2$ and is therefore
{\it positive}. 

Similarly, with 
the $\Delta^+$ wave function with maximal spin projection $M_J=3/2$ 
\begin{equation}
\vert \Delta^+  \rangle  =  {1 \over \sqrt{3} } \vert  
\left ( uud + udu  + duu \right ) \rangle 
\vert \uparrow \uparrow \uparrow \rangle ,
\end{equation}
we find for the intrinsic quadrupole moment of the $\Delta^+$ 
\begin{equation}
Q^{\Delta^+}_{0} = Q^{\Delta^+}= r_n^2.
\end{equation} 
In the case of the $\Delta$, there are no Clebsch-Gordan coefficients
that could be "renormalized," and there is no difference between the 
intrinsic $Q_0^{\Delta^+}$ and the spectroscopic quadrupole moment
$Q^{\Delta^+}$ 

Summarizing, in the quark model, 
the intrinsic quadrupole moment of the proton and the $\Delta^+$ 
are equal in magnitude but opposite in sign
\begin{equation}
\label{int2}
Q_0^p = - Q_0^{\Delta^+} .
\end{equation}
We conclude that the proton is a prolate and the 
$\Delta^+$ an oblate spheroid. The same conclusion is also
obtained in a quite different approach to which we turn in the next
section.

\subsection{Collective model}

Quadrupole moments of strongly deformed nuclei are not adequately described 
in a single-nucleon transition model. The measured quadrupole moments 
of strongly deformed nuclei exceed the quadrupole moment due
to a single valence nucleon in a deformed orbit usually by a  
factor of 10 or more. The collective nuclear model, which involves
the collective rotational motion of many nucleons of the nucleus,
gives a more realistic description of the data.

In the collective nuclear model \cite{Boh75}, the relation between the 
observable spectroscopic quadrupole moment $Q$ and the intrinsic quadrupole 
moment $Q_0$ is 
\begin{equation}
\label{collective}
Q= {3 K^2 -J(J+1) \over (J+1) (2J+3) } Q_0,
\end{equation}
where $J$ is the total spin of the nucleus,
and $K$ is the projection of $J$ onto the $z$-axis in the body fixed frame
(symmetry axis of the nucleus).
The intrinsic quadrupole moment $Q_0$ characterizes the deformation of the 
charge distribution in the ground state. The ratio between $Q_0$ and
$Q$ is the expectation value of the Legendre polynomial $P_2(\cos\Theta)$ 
in the substate with maximal projection $M=J$. This factor 
represents the averaging of the nonspherical charge distribtion due 
to its rotational motion as seen in the laboratory frame.
 
\begin{figure}[htb]
$$\mbox{
\epsfxsize 8.0 true cm
\epsfysize 10.5 true cm
\setbox0= \vbox{
\hbox{ 
\epsfbox{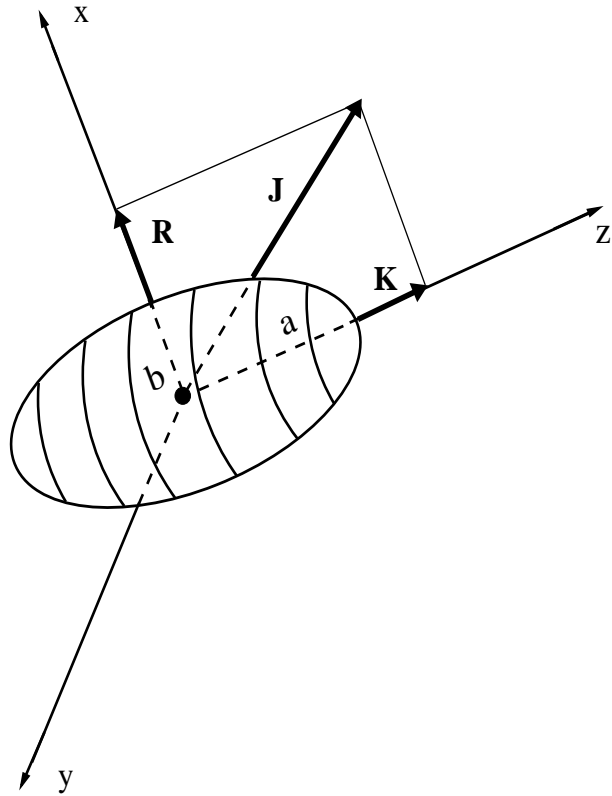}
} 
} 
\box0
} $$
\vspace{-0.4cm}
\caption[Collective]{Representation of the $\Delta$-isobar as a 
collective rotation of a prolate nucleon with  intrinsic spin $K=1/2$. 
The collective orbital angular momentum is denoted  by $R$.
As a result of the collective rotation of a cigar-shaped object ($N$)
with intrinsic spin $K=1/2$ one obtains a pancake-shaped object ($\Delta$) 
with total angular momentum  
$J=3/2$. The lengths of  major half-axis $a$ and minor half-axis $b$ 
can be calculated in the model of a homogeneously
charged spheroid (see text).}
\label{fig:collective}
\end{figure}

We consider the $\Delta$ with spin $J=3/2$ as a collective 
rotation of the entire nucleon with an intrinsic angular momentum $K=1/2$
(see Fig. \ref{fig:collective}).
This is how the $\Delta(1232)$ is viewed in the 
Skyrme model\footnote{With hindsight the Skyrme model is seen as
an effective field theory of QCD corresponding to the limit of infinitely 
many colors (and quarks)\cite{Bha88}.}. 
In this model,  the (rotational) energy of  
the $\Delta$ is inversely
proportional to its moment of inertia \cite{Bha88}. 
Therefore, it is energetically 
favorable to increase its moment of interia by assuming 
a deformed shape.
Inserting the quark model relation for the spectroscopic quadrupole moment 
$Q_{\Delta}= r_n^2$ on the left-hand side\footnote{This may be justified
because in the quark model $Q_{\Delta}$ 
is dominated by the collective $q \bar q $ degrees of freedom.} one finds 
for the intrinsic quadrupole moment of the proton
\begin{equation}
\label{int3} 
Q_0^p= - 5\,  r_n^2.  
\end{equation}

The large value for 
$Q_0^p$ is certainly due
to the crudeness of the model. The rigid rotor model for the nucleon 
which underlies Eq.(\ref{collective}) 
is most certainly an oversimplification. A more realistic 
description would treat nucleon rotation as being partly 
irrotational,  e.g., only the peripheral parts of the nucleon participate in
the collective rotation. This results in smaller intrinsic quadrupole 
moments\cite{Boh75}. However, we speculate that the sign of 
the intrinsic quadrupole moment  given by Eq.(\ref{int3}) is correct. 
If so, the nucleon is a prolate spheroid.

We can also use the collective model to estimate $Q_0^{\Delta}$.
For this purpose one regards the $\Delta^+$ as the 
$K=J=3/2$  ground state of a rotational band. We then obtain 
from Eq.(\ref{collective}) a negative intrinsic quadrupole moment 
for the $\Delta^+$
\begin{equation}
\label{int4} 
Q_0^{\Delta^+}= 5\,  r_n^2= -Q_0^p. 
\end{equation}
Obviously, the intrinsic quadrupole moments of the proton and the 
$\Delta^+$ have the same magnitude but different sign,
a result that was also obtained in the quark model with
two-quark operators. The sign change between $Q_0^p$ and $Q_0^{\Delta^+}$ 
can be explained by imagining a cigar-shaped ellipsoid ($N$) 
collectively rotating around the $x$ axis. This leads to a pancake-shaped 
ellipsoid ($\Delta$).
 
In classical electrodynamics the  simplest model for a nonspherical 
homogeneous charge distribution is 
a rotational ellipsoid with charge $Z$, major axis $a$ along, and minor 
axis $b$  perpendicular to the symmetry axis (see Fig. \ref{fig:collective}). 
Its quadrupole moment is given by 
\begin{equation}
\label{ellipsoid}
Q_0= {2 Z \over 5} (a^2-b^2) ={4 \over 5}\,  Z \, R^2 \, \delta,
\end{equation} 
with the deformation parameter $\delta=2(a-b)/(a+b)$ and 
the mean radius $R=(a+b)/2$. We use this model to estimate the 
degree of baryon deformation.

From the collective model we get  
$Q_0^p=0.565$ fm$^2$, and with the recent value for 
the proton charge 
radius  \cite{Ros00} we obtain for the `equivalent' radius 
$R_p =\sqrt{5/3}\,r_p=1.15$ fm. Eq.(\ref{ellipsoid}) then leads to
a deformation parameter 
$\delta_N \approx 0.53 $, and a  
ratio of major to minor semi-axes $a/b \approx 1.73$. 
Similarly, for the $\Delta$ one gets with the help of the charge radius
relation $r_{\Delta^+}^2 = r_p^2 - r_n^2$  \cite{Dil99,Leb00} 
$\delta_{\Delta}=-0.48$, and $a/b=-0.62$.
On the other hand, if we insert the quark model result 
$Q_0^p=-r_n^2=0.113$ fm$^2$ on the left-hand side of Eq.(\ref{ellipsoid}), 
we obtain a deformation parameter 
$\delta_N= 0.11 $. This corresponds  to 
a ratio of major to minor semi-axes $a/b= 1.11$.
For the deformation parameter of the $\Delta$ 
we find $\delta_{\Delta}=-0.09 $ and a half-axis ratio $a/b= -0.91$.

Summarizing, the collective model 
leads in combination with the experimental information 
to a positive intrinsic quadrupole moment of the nucleon 
and a negative intrinsic quadrupole moment for the $\Delta^+$. 
Although the magnitude of  the deformation is uncertain, 
we are confident that our assignment of a prolate deformation 
for the nucleon and an oblate deformation for the $\Delta$ is correct.

\subsection{Pion cloud model} 

Finally, we consider the physical proton with spin up, denoted by 
$\vert p \uparrow \rangle$, as
a coherent superposition of three different terms: (i) a 
spherical quark core contribution with spin 1/2, called a bare proton $p'$; 
(ii) a bare neutron $n'$ surrounded by a positively charged pion cloud; 
and (iii) a bare $p'$ surrounded by a neutral pion cloud
\cite{Hen62}. In the last two terms the spin(isospin) of the bare proton and 
of the pion cloud  are coupled to total spin and isospin of the physical 
proton.  Similarly, the physical $\Delta^+$ is considered as 
superposition of a spherical quark core term with spin 3/2, called a bare 
$\Delta^{+\, '}$, a bare $n'$ surrounded by a $\pi^+$ cloud, and  a bare 
$p'$ surrounded by a $\pi^0$ cloud. In each term, the spin/isospin  
of the quark core and pion cloud are coupled to the total spin and isospin 
of the physical $\Delta^+$. We then write:
\begin{eqnarray} 
\label{pionwave}
\vert p \uparrow \rangle &= & \alpha 
\vert p' \uparrow \rangle 
                         + \beta 
\frac{1}{3} \left (\vert p' \uparrow \pi^0 Y^1_0 \rangle 
-\sqrt{2} \vert p' \downarrow \pi^0 Y^1_1  \rangle 
-\sqrt{2} \vert n' \uparrow  \pi^+ Y^1_0  \rangle 
+ 2       \vert n' \downarrow  \pi^+ Y^1_1 \rangle \right )
\nonumber \\
\vert n \uparrow \rangle &= & \alpha 
\vert n' \uparrow \rangle 
                         + \beta 
\frac{1}{3} \left (- \vert n' \uparrow \pi^0 Y^1_0 \rangle 
+\sqrt{2} \vert n' \downarrow \pi^0 Y^1_1  \rangle 
+\sqrt{2} \vert p' \uparrow  \pi^- Y^1_0  \rangle 
- 2       \vert p' \downarrow  \pi^- Y^1_1 \rangle \right )
\nonumber \\
\vert \Delta^+ \uparrow \rangle &= & \alpha' 
\vert \Delta^{+'} \uparrow \rangle 
                         + \beta' 
\frac{1}{3} \left ( 2 \vert p' \uparrow \pi^0 Y^1_0 \rangle 
+ \sqrt{2} \vert p' \downarrow \pi^0 Y^1_1  \rangle 
+ \sqrt{2} \vert n' \uparrow  \pi^+ Y^1_0  \rangle 
+        \vert n' \downarrow  \pi^+ Y^1_1 \rangle \right ), \nonumber \\ 
&  & 
\end{eqnarray}
where $\beta$ and $\beta'$ describe
the amount of pion admixture in the $N$ and $\Delta$ wave 
functions. These amplitudes satisfy the normalization conditions 
$\alpha^2 + \beta^2=\alpha^{'2} + \beta^{'2} =1$, 
so that we have only two unknows $\beta$ and
$\beta'$. The $p$  and $\Delta^+$ wave functions are normalized 
and orthogonal.
Here, $Y^1_0$ and $Y^1_1$ are spherical harmonics of rank 1
describing the orbital wave functions of the pion. 
Because the pion moves predominantly in a $p$-wave,
the charge distributions of the proton and $\Delta^+$  
deviate from spherical symmetry, even if the bare proton and 
bare neutron wave functions are spherical.
 
The quadrupole operator to be used in connection with these states is
\begin{equation}
\label{pionquad}
{\hat Q_{\pi}} = e_{\pi} \sqrt{16 \pi \over 5} 
r_{\pi}^2 Y^2_0({\bf {\hat r}}_{\pi}),  
\end{equation}
where $e_{\pi}$ is the pion charge operator divided by the 
charge unit $e$, and $r_{\pi}$ is the distance between the center
of the quark core and the pion. Our choice of ${\hat Q}_{\pi}$ implies
that the quark core is spherical and the entire quadrupole moment 
comes from the pion p-wave orbital motion\footnote{A possible intrinsic 
deformation of the pion is neglected.}.

The $\pi^0$ terms 
do not contribute when evaluating the operator ${\hat Q}_{\pi}$ 
between the wave functions of Eq.(\ref{pionwave}).
We then obtain for the 
spectroscopic $\Delta$ and $N \to \Delta$  quadrupole moments 
\be
\label{pcm1}
Q_{\Delta^+}  = -{2 \over 15} \, {\beta'}^{2}\, r_{\pi}^2, \qquad 
Q_{p \to \Delta^+}  = {4 \over 15} \, {\beta'} \beta \, r_{\pi}^2. 
\ee
Only the $Y^1_1$ part of the pion wave function 
(pion cloud aligned in x-y plane) contributes to  $Q_{\Delta^+}$.
This leads to an oblate intrinsic deformation of 
the $\Delta^+$.

We have to determine three parameters $\beta$, $\beta'$, and $r_{\pi}$. 
From the experimental $N \to \Delta$ quadrupole transition moment,
$Q_{p \to \Delta}^{exp}\approx -0.11 =r_n^2 $ \cite{Bla97}, 
we can fix only one of them.
Therefore, we also calculate the 
nucleon and $\Delta$ charge radii in the pion cloud model and obtain
\bea 
\label{neutron}
r_p^2 & = & (1- \beta^2) r_{p'}^2 +\beta^2 
\left ( \frac{1}{3} r_{p'}^2  + \frac{2}{3} r_{\pi}^2 \right ) \nonumber \\
r_n^2 & = & \beta^2\left ( {2\over 3} r_{p'}^2 -{2\over 3} r_{\pi}^2 \right ) 
\nonumber \\
r_{\Delta^+}^2 & = & (1- {\beta'}^2) r_{p'}^2 +{\beta'}^2 
\left ( \frac{2}{3} r_{p'}^2  + \frac{1}{3} r_{\pi}^2 \right ).
\eea
Here, $r_{p'}^2$ is the charge radius of the bare proton. We 
assume that the charge radii of 
the bare proton and of the bare $\Delta^+$   
are equal. Adding the first two equations gives 
$r_{p'}^2= r_p^2 + r_n^2$, which expresses 
the bare proton charge radius in terms of
the experimental isoscalar nucleon charge radius. 
Subtracting the first and third equations one gets 
\be 
\label{cond}
r_p^2 - r_{\Delta^+}^2 = 
r_{p'}^2 \left ( \frac{1}{3} {\beta'}^2 
- \frac{2}{3} {\beta}^2 \right )  
+
r_{\pi }^2 \left ( \frac{2}{3} {\beta}^2 
- \frac{1}{3} {\beta'}^2 \right ) = r_n^2 = 
\beta^2\left ( {2\over 3} r_{p'}^2 -{2\over 3} r_{\pi}^2 \right ).
\ee
Because the correction to 
$r_p^2 - r_{\Delta^+}^2 = r_n^2 $ 
is of order $ {\cal O}(1/N_c^2)$ \cite{Dil99,Leb00} and therefore small, 
we obtain $\beta' =- 2\beta $.
When the latter condition is used in Eq.(\ref{pcm1}), we get  
\be 
Q_{\Delta^+}= Q_{p \to \Delta^+} = r_n^2.
\ee
This is 
in the same ballpark as the quark model prediction of Eq.(\ref{relations}).

We can now eliminate the model parameters and express them through
the experimental charge radii:
$\beta^2 = - (3/8){r_n^2}/({r_p^2+ r_n^2})$ 
and $r_{\pi}^2=  5 (r_p^2 + r_n^2) $.
The resulting  numerical values $\beta=0.26$, $\beta'=-0.52$,
$r_{\pi}=1.77$ fm correspond to a pion probability of $7\%$ in the
nucleon.  
The spatial extension of the pion cloud $r_{\pi}$ is close to the 
Compton wave length of the pion.
Due to the larger lever arm of $r_{\pi}$ compared to $r_{p'}$ 
the major part of the neutron charge radius and the nucleon's
intrinsic quadrupole moment comes from the pion cloud. 

\begin{figure}[htb]
$$\mbox{
\epsfxsize 12.0 true cm
\epsfysize 7.0 true cm
\setbox0= \vbox{
\hbox{ 
\epsfbox{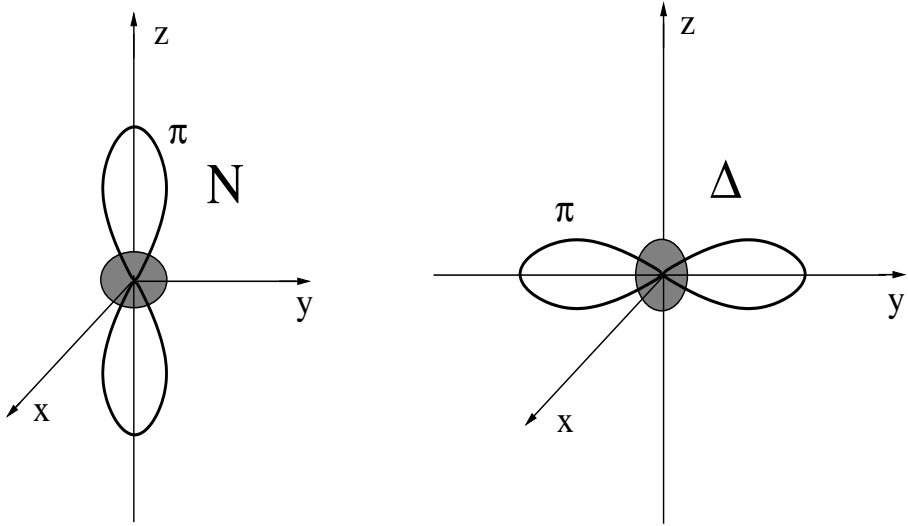}
} 
} 
\box0
} $$
\vspace{-0.3cm}
\caption[Pion cloud ]{Intrinsic quadrupole deformation of the nucleon (left)
and $\Delta$ (right) in the pion cloud model. In the $N$  
the $p$-wave pion cloud is concentrated along the polar (symmetry) axis, 
with maximum probability of finding the pion at the poles. 
This leads to a prolate deformation. In the $\Delta$, the pion cloud is 
concentrated in the equatorial plane producing an oblate intrinsic 
deformation. }
\label{fig:pcm}
\end{figure}

Next, we calculate the spectroscopic 
quadrupole moment of the proton in the pion
cloud model. We find  
\begin{equation} 
\label{pcm3}
Q^p=  {4 \over 3} \beta^2  r_{\pi}^2  \left (
{1 \over 3} \ \left ( { 2 \over 5 } \right ) 
+  {2 \over 3} \ \left ( -{ 1 \over 5} \right ) \right ). 
\end{equation} 
The factors $1/3$ and $2/3$ are the squares of the Clebsch-Gordan
coefficients that describe the angular momentum coupling of the 
bare neutron spin 1/2 with the pion orbital angular momentum $l=1$ to total
spin $J=1/2$ of the proton. They ensure that the spectroscopic
quadrupole moment of the proton is zero. The factors $2/5$ and 
$-1/5$ are the expectation values of the Legendre polynomial 
$P_2(\cos \Theta)$ evaluated between the pion wave function 
$Y^1_0({\bf {\hat r} }_{\pi})$ 
(pion cloud aligned along z-axis) and
$Y^1_1({\bf {\hat r}}_{\pi})$ (pion cloud aligned along an axis 
in the x-y plane). 
If we set {\it by hand} each of the coupling coefficients in front of 
$<Y^1_0| P_2| Y^1_0> $ and $<Y^1_1| P_2| Y^1_1> $  equal to 1/2,
the cancellation between the two orientations of the cloud  disappears.
The normalization of the sum of coupling coefficients is thereby 
preserved.  We note that the first term in Eq.(\ref{pcm3}), 
which comes from the $Y^1_0$ part of the pion wave function, dominates. 
Therefore, the probability for finding the pion in the nucleon 
is largest at the poles. This term is just 
the negative  of the spectroscopic $\Delta^+$ quadrupole moment.

By this procedure 
we are undoing the geometric averaging over all angles, 
which prevents the nonsphericity of the  pion cloud from being observed  
in the laboratory. One then finds for the intrinsic quadrupole moment of 
the proton and the $\Delta^+$  
\begin{equation}
\label{pcm4}
Q^p_0=  {4 \over 3} \beta^2  r_{\pi}^2  \left (
{1 \over 2} \ \left ( { 2 \over 5 } \right ) 
+  {1 \over 2} \ \left ( { 2 \over 5} \right ) \right )=  
{8 \over 15} \beta^2 r_{\pi}^2 = -r_{n}^2,  \qquad 
Q^{\Delta^+}_0 = r_n^2. 
\end{equation} 
Again, the intrinsic quadrupole moment of the $p$  is positive 
and that of the $\Delta^+$ negative. They are
identical in magnitude but opposite in sign.

The positive sign of the intrinsic proton\footnote{
After dividing by the negative
sign of the $\pi^-$ cloud, the neutron's
intrinsic quadrupole moment is also positive, i.e., $Q_0^n=Q_0^p$.}.  
quadrupole moment 
has a simple geometrical interpretation in this model.
It arises because the pion is preferably
emitted along the spin (z-axis) of the nucleon (see Fig. \ref{fig:pcm}). 
Thus, the proton assumes a prolate shape.
Here, we neglect the deformation of the bare nucleon
quark bag due to the pressure of the surrounding pion cloud.  
We emphasize that in this model 
all of the deformation comes from the pion cloud
itself, none from the valence quark core.
Previous investigations in a quark model with pion exchange \cite{Ven81} 
concluded 
that the nucleon assumes an oblate shape under the pressure of the
surrounding pion cloud, which is strongest along the polar axis.
However, in  these studies the deformed shape of the pion cloud 
itself was ignored. Inclusion of the latter  
leads to a prolate deformation that exceeds the small
oblate quark bag deformation by a large factor. 

\section{Summary} 

The experimental evidence for nonvanishing 
$N\to \Delta$ and $\Delta$ quadrupole  
moments can be seen as an indication for an intrinsic nucleon deformation.
In the present paper, the intrinsic quadrupole moment of the nucleon has 
been estimated in (i) a quark model, (ii) a collective model, and
(iii) a pion cloud model, using the empirical 
information on the $p \to \Delta^+$ quadrupole moment and the nucleon
charge radii.

The quark model  with the two-body quadrupole operator, 
and the pion cloud model  
predict a negative sign for the 
{\it spectroscopic} $\Delta^+$ quadrupole moment, and  that 
$Q_{\Delta^+} \approx Q_{p \to \Delta^+} \approx r_n^2$.
In all three models we find that the 
{\it intrinsic} quadrupole moment
of the nucleon is {\it positive}. This indicates a prolate shape of the
proton charge distribution. On the  other hand, 
the intrinsic quadrupole moment
of the $\Delta^+$ is found to be   
{\it negative}, indicative of an oblate $\Delta^+$
charge distribution. 
As to the magnitude of the deformation, 
the models vary within a  wide range $Q_0^p=0.11-0.55$ fm$^2$.

Despite their differences, all models 
emphasize collective over single-particle degrees of freedom and 
lead to an appreciable spectroscopic
$p \to \Delta^+$ transition quadrupole moment,
in agreement with recent experiments.
In our opinion this reflects that the intrinsic nucleon 
deformation resides 
mainly in the $q \bar q$  cloud surrounding an almost spherical 
valence quark core. It would be interesting to calculate the intrinsic 
quadrupole moment of the nucleon in other models, in order to check whether 
our finding of a prolate nucleon shape can be confirmed.

\vspace{0.5 cm}

{\bf Acknowledegement}: E. M. H. thanks Professor Amand Faessler and the 
members of the Institute for Theoretical Physics at the University
of T\"ubingen for hospitality during his stay,
when the idea for this work was conceived. He also thanks the Alexander
von Humboldt Foundation for a grant. A. J. B. thanks 
the Nuclear Theory Group of the University of Washington 
for hospitality and some financial support.


\begin{thebibliography}{99}
\bibitem{Ros00} R. Rosenfelder, Phys. Lett. {\bf B479}, 381 (2000).
\bibitem{Boh75} A. Bohr and B. Mottelson, Nuclear Structure II,
W. A. Benjamin, Reading (1975); J. M. Eisenberg and W. Greiner, 
Nuclear Models, North Holland, Amsterdam (1970),   see also 
P. Brix, Z. Naturforsch. {\bf 41a}, 3 (1986); 
P. Brix und H. Kopfermann, Z. Phys. {\bf 126}, 344 (1949). 
\bibitem{Bec65} C. Becchi and G. Morpurgo, Phys. Lett. {\bf 17}, 352 (1965).
\bibitem{Bec97} R. Beck, H. P. Krahn {\it et al.}, Phys. Rev. Lett. {\bf 78},
606 (1997).
\bibitem{Bla97} 
 G. Blanpied {\it et al.} Phys. Rev. Lett. {\bf 79}, 4337 
(1997); Phys. Rev. {\bf C} (2000) submitted, BNL-67526 preprint.
\bibitem{Kal97} F. Kalleicher {\it et al.} Z. Phys. {\bf A359}, 201 (1997). 
\bibitem{Bar98} for more recent measurements of $E2/M1$ and $C2/M1$ 
and theoretical developments see: Proceedings of Baryons '98, Bonn Germany,
World Scientific, Singapore, 1999, eds. D. W. Menze and B. Metsch.
\bibitem{Gia79} M. M. Giannini, D. Drechsel, H. Arenh\"ovel, and 
V. Tornow, Phys. Lett. {\bf B88}, 13 (1979).
\bibitem{Ven81} V. Vento, G. Baym, and A. D. Jackson, 
Phys. Lett. {\bf B102}, 97 (1981).
\bibitem{Ma83} Z. Y. Ma and J. Wambach, Phys. Lett. {\bf B132}, 1 (1983).
\bibitem{Cle84} G. Cl\'ement and M. Maamache,
Ann. of Physics (N.Y.) {\bf 165}, 1 (1984).
\bibitem{Mig87} A. B. Migdal, JETP Lett. {\bf 46}, 322 (1987).
\bibitem{Buc99} A. J. Buchmann, in Baryons '98, 
World Scientific, 1999, eds. D. W. Menze and B. Metsch, pg. 731.
\bibitem{Ger82} S. S. Gershtein and G. V. Dzhikiya,
Sov. J. Nuc. Phys. {\bf 34}, 870 (1982).
\bibitem{Isg82} N. Isgur, G. Karl and R. Koniuk, 
Phys. Rev. {\bf D25}, 2394 (1982).
\bibitem{Buc91}
A. J. Buchmann, E. Hern\'andez, and K. Yazaki, Phys. Lett.
{\bf B269}, 35 (1991);   Nucl. Phys. {\bf A569}, 661 (1994). 
\bibitem{Buc00} A. J. Buchmann and E. M. Henley, unpublished 
\bibitem{Gia90}
M. M. Giannini, Rep. Prog. Phys. {\bf 54}, 453 (1990). 
Note that we omit the unit charge  $e$ here and in the following
and give the quadrupole moments in units of fm$^2$.
\bibitem{Buc97} A. J. Buchmann, E. Hern\'andez, and Amand Faessler,
Phys. Rev. {\bf C55}, 448 (1997);  A. J. Buchmann, E. Hern\'andez, U. Meyer, 
and Amand Faessler, Phys. Rev. {\bf C58}, 2478 (1998).
\bibitem{Cap90} S. Capstick and G. Karl, Phys. Rev. {\bf D41}, 2767 (1990).
\bibitem{Mor89} G. Morpurgo, Phys. Rev. {\bf D40}, 2997 (1989).  
\bibitem{Kum72} K. Kumar, Phys. Rev. Lett. {\bf 28}, 249 (1972).
\bibitem{Bha88} R. K. Bhaduri, Models of the nucleon, Lecture notes and 
supplements in physics: 22, ed. Davis Pines, Addison-Wesley,
Redwood City, California, 1988.
\bibitem{Hen62} E. M. Henley and W. Thirring, Elementary Quantum Field
Theory, McGraw-Hill, New-York, 1962.
\bibitem{Dil99} G. Dillon and G. Morpurgo, Phys. Lett. {\bf B448}, 107 (1999). 
\bibitem{Leb00} A. J. Buchmann and R. F. Lebed, 
Phys. Rev.{\bf D62}, 096005 (2000). 


\end{thebibliography}
\end{document}